\documentclass[conference]{IEEEtran}
\IEEEoverridecommandlockouts
\usepackage{threeparttable}
\usepackage{cite}
\usepackage{amsthm,amsmath,amssymb,amsfonts}
\usepackage{algorithmic}
\usepackage{graphicx}
\usepackage{textcomp}
\def\BibTeX{{\rm B\kern-.05em{\sc i\kern-.025em b}\kern-.08em
    T\kern-.1667em\lower.7ex\hbox{E}\kern-.125emX}}
\usepackage{array}
\usepackage{booktabs}
\usepackage{multirow}
\usepackage[export]{adjustbox}

\usepackage[T1]{fontenc}
\usepackage[utf8]{inputenc}
\usepackage{lmodern}
\usepackage{hyperref}
\hypersetup{draft}

\makeatletter
\def\ps@IEEEtitlepagestyle{%
	\def\@oddfoot{\mycopyrightnotice}%
}
\def\mycopyrightnotice{%
	{\small 978-1-5386-4184-2/18/\$31.00~\copyright~2018 IEEE \hfill}
}

\newcommand{\dm}{{DroidMorph}}

\begin{document}\sloppy

\title{DroidMorph: Are We Ready to Stop the  Attack of Android Malware Clones?}

\author{\IEEEauthorblockN{Shahid Alam}
\IEEEauthorblockA{\textit{Dept. of Computer Engineering} \\
\textit{Adana Science and Technology University}\\
Adana, Turkey \\
salam@adanabtu.edu.tr}
\and
\IEEEauthorblockN{M. Zain ul Abideen and Shahzad Saleem}
\IEEEauthorblockA{\textit{Dept. of Computing} \\
\textit{National University of Science and Technology}\\
Islamabad, Pakistan \\
\{mabideen.msis18seecs,shahzad.saleem\}@seecs.edu.pk}
}

\maketitle

\begin{abstract}

The number of Android malware variants (clones) are on the rise and, to stop this attack of clones we need to develop new methods and techniques for analysing and detecting them. As a first step, we need to study how these malware clones are generated. This will help us better anticipate and recognize these clones. In this paper we present a new tool named DroidMorph, that provides morphing of Android applications (APKs) at different level of abstractions, and can be used to create Android application (malware/benign) clones. As a case study we perform testing and evaluating resilience of current commercial anti-malware products against attack of the Android malware clones generated by DroidMorph. We found that 8 out of 17 leading commercial anti-malware programs were not able to detect any of the morphed APKs. We hope that DroidMorph will be used in future research, to improve Android malware clones analysis and detection, and help stop them.
\end{abstract}

\begin{IEEEkeywords}
Android APK, Malware variant, Morphing, Obfuscation.

\end{IEEEkeywords}

\section{Introduction}

According to the recent Symantec threat reports, Android continues to be the most targeted mobile platform, the number of new mobile malware attacks grew by 105\% from 2015 to 2016 \cite{Symantec-2017}, and the number of new discovered mobile malware variants (clones) grew by 54\% from 2016 to 2017 \cite{Symantec-2018}. In addition to these simple attacks of \emph{clones}, there are also Android malware \emph{clones of clones}, i.e., \emph{clones} of a malware family which themselves are \emph{clones}. For example, \emph{DroidKungFu1}, \emph{DroidKungFu2}, \emph{DroidKungFu3} and \emph{DroidKungFu4} are 4 different families of the original Android \emph{DroidKungFu} malware, and each of these 4 families have their own \emph{clones} (variants) \cite{Dissecting-AM}. According to the McaFee threat report, the number of malware families found in the Google play increased by 30\% in 2017 \cite{McAfee-threat-report-2017}.

Malware writers use stealthy mutations (\emph{morphing/obfuscations}) to continuously develop malware clones, thwarting detection by signature based detectors. This attack of clones seriously threatens all the mobile platforms, especially Android. In the rest of the paper we use variants and clones interchangeably.

As mentioned before, the number of Android malware clones are on the rise and, to stop this attack we need to develop new methods and techniques for analysing and detecting them. As a first step, we need to study how these malware clones are generated. This will help us better anticipate and recognize these clones. This is the main motivation for the research carried out in this paper.

In this paper, we present the design and development of an Android APK morphing tool, and using this tool evaluate resilience of the current commercial antimalware products against attack of the Android malware clones. This work also complements other such previous works \cite{Obfuscation-Faruki, Droidchameleon-journal, ADAM, Pandora-2013}.

The contributions of this work are as follows:

\begin{itemize}
	\item Designing and developing of a morphing tool named \emph{\dm}, that provides morphing of Android applications (APKs) at different level of abstractions, and can be used to create Android application (malware/benign) clones.

	\item Testing and evaluating resilience of current commercial antimalware products against attack of the Android malware clones generated by \emph{\dm}. We found that 8 out of 17 leading commercial anti-malware programs were not able to detect any of the morphed APKs.
\end{itemize}

The remainder of this paper is organized as follows. We present related work in Section \ref{sec:related_works}. A brief overview of \emph{\dm} is presented in Section \ref{sec:design}. Section \ref{sec:evaluation} presents the evaluation of \emph{\dm} by testing current commercial antimalware products against attack of the Android malware clones generated by \emph{\dm}. Section \ref{sec:conclusion} finally concludes the paper.

\section{Related Work and Comparison}\label{sec:related_works}

In this Section, we briefly compare four of the previous works that have evaluated the resilience of commercial antimalware products with \emph{\dm}.

Faruki et al. \cite{Obfuscation-Faruki} evaluated anti malware products against control, data and layout transformations. Anti-malware malware programs were then evaluated against different permutations of these three transformations. They found that top rated anti-malware programs are vulnerable against permutation of these transformations.

Rastogi et al. \cite{Droidchameleon-journal} evaluated anti malware products against a combination of trivial and non-trivial obfuscations. They tested 10 anti-malware programs. Repetitive transformations were used to fail the anti-malware program.

Zheng et al. \cite{ADAM} evaluated anti malware products against some of the non-trivial obfuscations. They collected 222 Android malware samples from the wild. These samples were then transformed to test them on VirusTotal \cite{VirusTotal} against different anti-malware programs. They reported that one of the anti-malware program tested (AntiY AVL) is better than the others.

Protsenko et al. \cite{Pandora-2013} evaluated anti malware products against data and object-oriented design obfuscations. They tested 10 popular anti-malware programs and found deficiencies in most of them. They also compared and presented the shortcomings of a state of the art static similarity tool.

All these four works have implemented prototype tools to create variants (morph) of known malware to test and evaluate commercial antimalware products. Here are some of the differences between \emph{\dm} and the other four works.

\begin{itemize}
	\item
	\emph{\dm} is implemented on top of the Soot Framework \cite{SOOT-1999}, whereas \cite{Obfuscation-Faruki, Droidchameleon-journal, ADAM} use third party tools for assembling and disassembling, and another set of tools for morphing. This dependency on a set of third party tools may make their \cite{Obfuscation-Faruki, Droidchameleon-journal, ADAM} implementation unstable and inflexible, and incompatible to work with different Android OS APKs. Soot provides all necessary functionality regarding analysis, modification, and generation of Android/Java bytecode and is compatible with different Android OS APKs.

	\item
	Unlike \cite{Pandora-2013}, \emph{\dm} provides direct support for analysis, modification, and generation of Android (Dalvik) bytecode.

	\item
	Beside providing non-trivial and trivial obfuscations \emph{\dm} also provides morphing of Android APK at different level of abstractions.
\end{itemize}

\section{Brief Overview of \dm}\label{sec:design}

\begin{figure}[htb]
	\centering
	\includegraphics[scale=0.45]{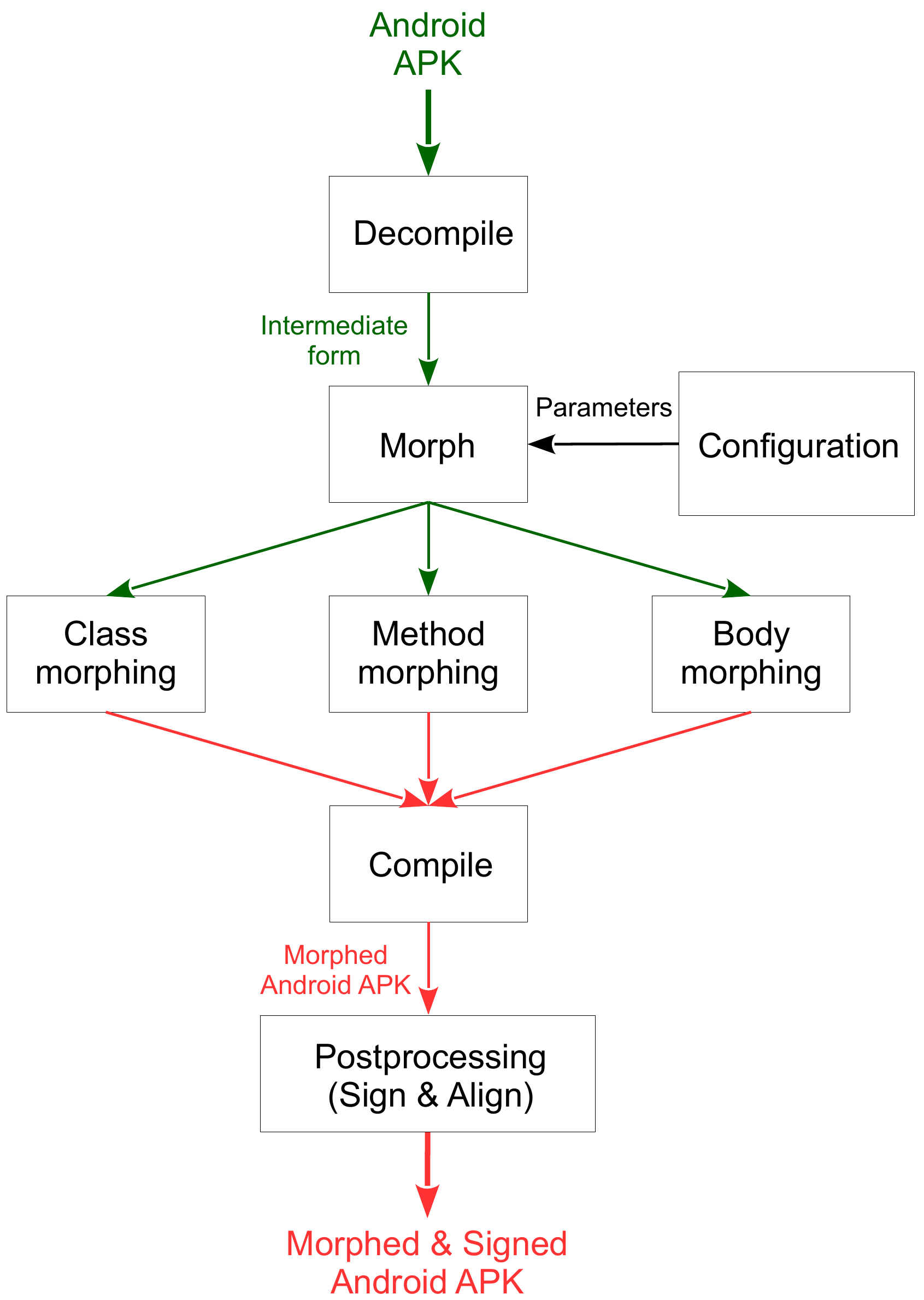}
	\caption{Architectural Overview of \dm}
	\label{fig:DroidMorph}
\end{figure}

Figure \ref{fig:DroidMorph} provides a high level architectural overview of \emph{\dm}. We first decompile an Android APK to an intermediate form. Morphing is carried out at different level of abstractions on this intermediate form. The morphed intermediate form is then compiled to morphed Android APK. This APK is then signed to generate the final morphed and signed Android APK that is ready to run on the Android platform.

Current design of \emph{\dm} implements morphing at three level of abstractions, \emph{Class}, \emph{Method} and \emph{Body}. The implementation of trivial and non-trivial obfuscations is work in progress. So far as trivial obfuscations, we have implemented morphing of \emph{Class} and \emph{Method} names, and as non-trivial obfuscations, we have implemented morphing of \emph{Control Flow Graph} and \emph{Call Flow Graph}.

\begin{table}[ht]
	\caption{Class distribution of the 848 Android malware samples}
	\setlength{\tabcolsep}{12.5pt}
	\renewcommand{\arraystretch}{2.0}
	\centering
	\begin{tabular}{ l c c } \hline\hline
		Class/family 			&			Number of samples 			\\ \hline
		AnserverBot 			&					187	 				\\
		BaseBridge 				&					122					\\
		DroidKungFu3 			&					309					\\
		DroidKungFu4 			&					96 					\\
		DroidDream 				&					16 					\\
		DroidDreamLight 		&					46 					\\
		Geinimi 				&					69 					\\ \hline
	\end{tabular}
	\label{table:dataset_malware}
\end{table}

\section{Experimental Evaluation}\label{sec:evaluation}

We carried out an empirical study to analyse the correctness and efficiency of \emph{\dm}. We present in this section the empirical study, obtained results and analysis.


\subsection{Dataset}

Our dataset for the experiments consists of 848 Android malware programs collected from two different resources \cite{Dissecting-AM, Contagiominidump}. Table \ref{table:dataset_malware} shows distribution of these malware samples. The malware dataset shows a variety of samples from different families.

\begin{table*}[ht]
	\caption{Distribution of 1771 malware variants generated by \emph{\dm}.}
	\setlength{\tabcolsep}{18pt}
	\renewcommand{\arraystretch}{1.85}
	\centering
	\begin{tabular}{  l  c  c  c  c  } \hline\hline
		
		\textbf{Class/family} 	& \multicolumn{4}{  c  }{\textbf{Number of variants generated for each level of abstraction}} \\
								& Class morphing & Method morphing & Body morphing & All morphing \\ \hline

		AnserverBot 			&	2	&	182	&	171	& 	2			\\
		BaseBridge 				&	1	&	117	&	73 	&	1			\\
		DroidKungFu3 			&	63	&	291	&	281	&	59			\\
		DroidKungFu4 			&	21	&	91	&	95 	&	19			\\
		DroidDream 				&	0	&	16	&	15 	&	0			\\
		DroidDreamLight 		&	0	&	44	&	44 	&	0			\\
		Geinimi 				&	3	&	66	&	48 	&	3			\\ \hline
	\end{tabular}
	\label{table:malware-variants}
\end{table*}

\begin{table*}[ht]
	
	\caption{Detection results of the 17 commercial anti-malware programs tested with 1771 variants of 7 malware families generated by \emph{\dm} listed in Table \ref{table:malware-variants}}.
	\setlength{\tabcolsep}{18pt}
	\renewcommand{\arraystretch}{1.5}
	\centering
	\begin{tabular}{  l  c  c  c  c  } \hline\hline
		
		\textbf{Anti-malware} 	& \multicolumn{4}{  c  }{\textbf{Detection Rate (\%)}} \\
		& Class morphing & Method morphing & Body morphing & All morphing \\ \hline

		AVG & 95 &  100 &  100 &  100    \\ \hline
		BitDefender &  100 &  100 &  100 &  100    \\ \hline
		LineSecurity &  0 &  0 &  0 &  0    \\ \hline
		Kaspersky &  100 &  100 &  100 &  100    \\ \hline
		Sophos &  100 &  100 &  100 &  100    \\ \hline
		MaxSecurity &  0 &  100 &  8 &  0    \\ \hline
		DrWeb &  95 &  100 &  95 &  100    \\ \hline
		ESET &  83.3 &  100 &  100 &  100    \\ \hline
		DUSecurityLabs &  0 &  0 &  0 &  0    \\ \hline
		VIPRE &  100 &  100 &  100 &  100    \\ \hline
		AntivirusPro &  0 &  0 &  17 &  0    \\ \hline
		360Security &  0 &  0 &  0 &  0    \\ \hline
		McAfee &  100 &  100 &  100 &  100    \\ \hline
		SecuritySystems &  0 &  0 &  0 &  0    \\ \hline
		GoSecurity &  0 &  0 &  0 &  0    \\ \hline
		LAAntivirusLab &  0 &  0 &  0 &  0    \\ \hline
		Malwarebytes & 100 &  100 &  100 &  100    \\ \hline
		\textbf{Average Detection Rate} &  51.4 &  58.8 &  54.1 &  52.9    \\ \hline
	\end{tabular}
	\label{table:results-variants}
\end{table*}

\subsection{Evaluation}

We used \emph{\dm} to generate variants of each malware family listed in Table \ref{table:dataset_malware}. The distribution of these generated malware variants are shown in Table \ref{table:malware-variants}. Randomly selected of these malware variants were tested for there correctness by running them on an Android phone.

We used VirusTotal \cite{VirusTotal} and selected 17 anti-malware programs for our empirical study. These anti-malware programs were then used to detect the malware variants generated by DroidMorph listed in Table \ref{table:malware-variants}. The results are shown in Table \ref{table:results-variants}. The results show that 8 out of 17 anti-malware programs were not able to detect any morphed APKs. Different anti-malware programs use different detection techniques like signature based, static based and behavior based detection, which use different signature and/or anomaly databases.

The result shows that DroidMorph was successful to bypass the security of 8 anti-malware programs used. The average detection rate of class morphing was better than all morphing. It is because class contains (Table \ref{table:malware-variants}) more variants than all morphing. It is possible that anti-malware programs were able to detect less variants of DroidKungFu3 and DroidKungFu4 in class than all. To make an exact comparison we need to use equal number of variants from the same families. Anyhow, this needs more investigation. Research on DroidMorph is a work in progress, in future more work needs to be done on class level morphing which will also increase the morphing on all level increasing the effectiveness of DroidMorph.

\section{Conclusion}\label{sec:conclusion}

The number of Android malware clones are on the rise and to stop this attack of clones we need to study how these clones are generated. As a first step, in this paper we have presented a new tool named DroidMorph, that provides morphing of Android applications (APKs) at different level of abstractions. We used DroidMorph to generate Android malware clones to evaluate commercial anti-malware products and found that 8 out of 17 leading commercial anti-malware programs were not able to detect morphed APKs successfully. We hope that DroidMorph will be used in future research, to improve Android malware clones analysis and detection, and help stop them.

In DroidMorph, we have only implemented some basic trivial and non-trivial obfuscations (morphing). Implementation of other obfuscations is work in progress, which will further improve (reduce detection by anti-malware programs) the results. In future, we will further improve morphing at different levels, specifically class level morphing. We will also add morphing of meta information (permissions etc.) embedded in an APK which will further reduce the detection by anti-malware programs.

\bibliographystyle{IEEEtran}
\bibliography{MobileMalware}

\end{document}